\documentclass[9pt,twocolumn]{extarticle}
\pdfoutput=1

\usepackage[sort&compress]{natbib}
\setcitestyle{super}
\setcitestyle{square} 
\setcitestyle{comma} 
\setcitestyle{open={}}
\setcitestyle{close={}}
\usepackage{graphicx}
\usepackage[margin=1in]{geometry}
\usepackage[usenames,dvipsnames]{color}
\usepackage{amsmath,amssymb}
\usepackage{multicol}
\usepackage{mathpazo} 
\usepackage{courier} 
\normalfont
\usepackage[T1]{fontenc}
\usepackage{caption}
\newcommand{\chithree}{ \ensuremath{\chi^{(3)}} }
\newcommand{\ket}[1]{\ensuremath{|#1\rangle\mkern-1mu}}
\newcommand{\bra}[1]{\ensuremath{\mkern-1mu\langle#1|}}

\newcommand{\ad}[1]{\textsuperscript{#1}\kern-2pt}

\newcommand{\pseudosection}[1]{\vspace{9pt}\noindent\textbf{#1}}

\def\({\left(}
\def\){\right)}
\def\[{\left[}
\def\]{\right]}

\usepackage{capt-of}
\newcommand{\makefig}[4][189mm]{
\begingroup\makebox[\textwidth][c]{\includegraphics[width=#1]{"#2"}}
\captionof{figure}{\textbf{\normalsize|} \textsf{#3}}\label{#4}\endgroup
}
\newcommand{\makefigc}[4][88mm]{
\begin{figure}[tb] \begin{center}
\includegraphics[width=#1]{"#2"}
\caption{\textbf{\normalsize|} \textsf{#3}}\label{#4}
\end{center}\end{figure}}

\def\thSY{\theta_\mathrm{SY}}
\def\thSZ{\theta_\mathrm{SZ}}
\def\thIY{\theta_\mathrm{IY}}
\def\thIZ{\theta_\mathrm{IZ}}
\def\balance{\beta}
\def\overlap{\sigma}
\def\Ry{\hat R_y}
\def\Rz{\hat R_z}
\def\theorystate{{\hat\rho}_\text{th}}
\def\expstate{{\hat\rho}_\text{ex}}

\def\mytitle{Qubit entanglement on a silicon photonic chip}

\title{\vspace{-1.0cm}\Huge\textbf{\textsf{\mytitle}}}
\author{\textsf{J. W. Silverstone\ad{1*}, R. Santagati\ad{1*}, D. Bonneau\ad{1}, M. J. Strain\ad{2}, M. Sorel\ad{3}, J. L. O'Brien\ad{1}, M. G. Thompson\ad{1}}}
\date{}

\begin{document}
\twocolumn[{%
\maketitle
\vspace{-5mm}
\begin{center}
\begin{minipage}{0.7\textwidth}
\begin{center}
\textsf{\footnotesize\textsuperscript{1}Centre for Quantum Photonics, H. H. Wills Physics Laboratory and Department of Electrical and Electronic Engineering, University of Bristol, Merchant Venturers Building, Woodland Road, Bristol BS8 1UB, UK. \\\textsuperscript{2} Institute of Photonics, Department of Physics, University of Strathclyde,\\ Wolfson Centre, 106 Rottenrow East, Glasgow G4 0NW, UK. \\\textsuperscript{3} School of Engineering, University of Glasgow, James Watt South Building, Glasgow G12 8QQ, UK. \\\textsuperscript{*} Authors J.W.S. and R.S. contributed equally to this work.\\}
\end{center}
\end{minipage}
\vspace{+4mm}
\end{center}

\setlength\parindent{12pt}

\begin{multicols}{2}

\noindent
\textbf{Entanglement---one of the most delicate phenomena in nature---is an essential resource for quantum information applications\cite{Raussendorf2003, Horodecki2009}. Large entangled cluster states have been predicted to enable universal quantum computation\cite{Briegel2009}, with the required single-qubit measurements readily implemented with photons\cite{OBrien2009, Nielsen2004}. Useful large-scale systems must generate and control qubit entanglement on-chip, where quantum information is naturally encoded in photon path. Here we report a silicon photonic chip which integrates resonant-enhanced sources, filters, and reconfigurable optics to generate a path-entangled two-qubit state---the smallest non-trivial cluster state---and analyse its entanglement. We show that ring-resonator-based spontaneous four-wave mixing sources can be made highly indistinguishable, despite their nonlinear dynamics\cite{Pernice2011, Pernice2010, Priem2005}, and the first evidence that their frequency correlations are small, as predicted\cite{Helt2010}. We use quantum state tomography, and the strict Bell-CHSH inequality\cite{Aspect1982} to quantify entanglement in the device, confirming its high performance. This work integrates essential components for building devices and systems to harness quantum entanglement on the large scale.}

Quantum entanglement is at the heart of quantum information science: entanglement between photons and the vacuum gives security to quantum communications channels; entanglement between photons passing through a sample enables its super-resolution measurement; and entanglement between qubits provides the tremendous power behind quantum computation. Entanglement is regularly generated in bulk- or fibre-based quantum optical systems by directly using the intrinsic polarisation correlations of the photon-pair source\cite{Yao2012, Bell2013}, and on-chip using post-selected logic gates\cite{Shadbolt2012, Politi2008}. On-chip sources of photon pairs have been recently developed\cite{Silverstone2014, Collins2013, Sharping2006, Agrawal2006}, but rely on nonlinear processes in which all fields---pump, signal, and idler---are co-polarised, both due to the increased strength of such processes, and due to the difficulty of controlling polarisation with integrated optics (with some exceptions \cite{Matsuda2012, Crespi2011, Olislager2013}). Since source-based entanglement typically lies in the photonic polarisation degree of freedom, on-chip sources of path qubit entanglement have been scarce.

We present a silicon-on-insulator photonic chip, operating in the central telecommunications band, which can generate and analyse the path entanglement produced by two coherently pumped photon-pair sources. As shown in Fig.~\ref{fig:schematic}, a pulsed pump laser is launched into two microring photon-pair sources which produce pairs in a superposition between being created in one source or the other. The device reconfigures this superposition, using on-chip filters and a waveguide crossing, into an entanglement between two photonic path qubits. Finally, these path qubits are analysed using two on-chip Mach-Zehnder interferometers. The pump laser, pump-suppressing filters, and detectors are all fibre-integrated, off the chip. In this letter, we integrate narrow-band photon sources, spectral filters, and reconfigurable photonic systems into a single device. We describe and quantify the performance of each of these functionalities, culminating with a precise estimation of the on-chip path-entangled state, and a strict measurement of its entanglement.

Spontaneous four-wave mixing (SFWM)\cite{Sharping2006, Agrawal2006} is an effect of the third order nonlinear susceptibility \chithree of the medium---the silicon waveguide core. We use SFWM to produce photon pairs on-chip. By convention, the two constituents of each pair are referred to as `signal' and `idler' photons, with frequencies $\nu_s$ and $\nu_i$, equally spaced on either side of the pump frequency $\nu_p$; we will refer to the higher-energy photon as the signal (i.e. $\nu_s > \nu_p > \nu_i$).
\end{multicols}

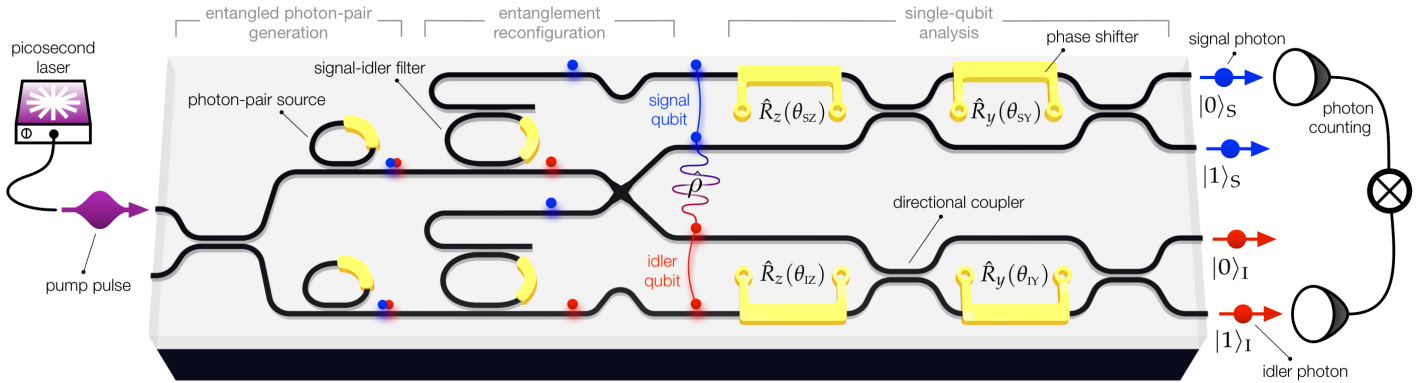
\makefig{fschem}{Schematic layout of the device. A picosecond pump pulse is coupled into the silicon chip where it generates a superposition of photon pairs via spontaneous four-wave mixing. This superposition is separated into signal (blue) and idler (red) path qubits, which are analysed by two Mach-Zehnder interferometers. Photons at the output are separated from residual pump by fibre wavelength-division multiplexers (not shown) and collected by single-photon detectors.}{fig:schematic}

}]

\noindent  
SFWM acts to annihilate two photons from the (degenerate) pump and create the signal and idler at new frequencies, via the phenomenological Hamiltonian $\hat H \propto a_i^\dagger a_p^2 a_s^\dagger + a_i {a_p^{\dagger}}^2 a_s$. Each SFWM event conserves the energy and momentum of the input photons. In our experiment, SFWM occurs in the optical cavities formed by two microring resonators, which modify the density of states of the parametric fluorescence, and structure the spectrum of these photon pairs into bright fluorescent peaks around the cavity resonances\cite{Azzini2012}. This structure differs from the characteristic flat, broad spectrum of straight-waveguide-based sources, which is shaped by momentum conservation alone.

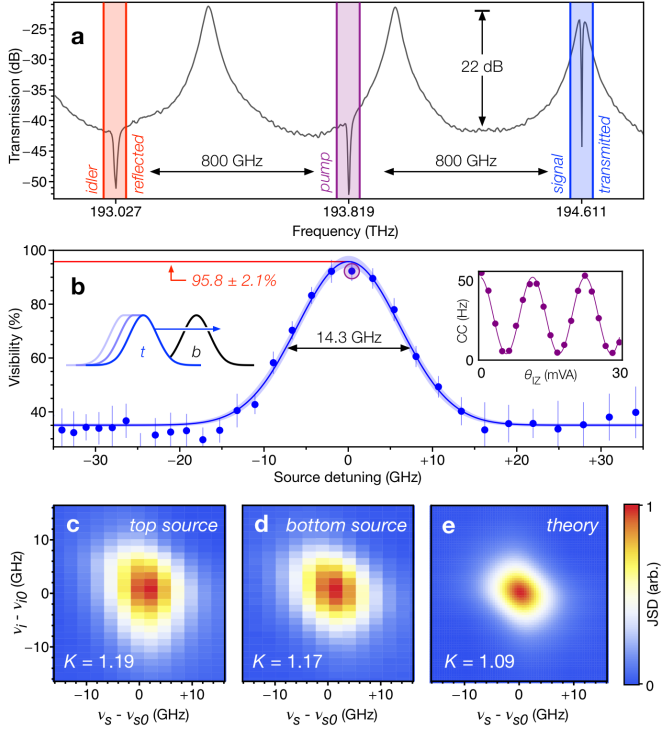
\makefigc{fspect}{Spectral characteristics of the experiment. \textbf{a,} Spectral layout of source (dips) and filter (peaks) resonances in the central telecommunications band. Source free spectral range is 800~GHz, to match the 200~GHz ITU grid. \textbf{b,} Two-photon fringe visibility measured as a function of top-to-bottom source detuning, as the top resonances were scanned over the stationary bottom resonances. Inset: representative two-photon fringe corresponding to peak visibility value (highlighted). The residual visibility is due to interference with pairs generated in the straight waveguide. Error bars represent three standard errors of each sinusoidal regression. Shaded region on fit represents one standard deviation in visibility. Measured joint spectral density profiles for the top (\textbf{c}), and bottom (\textbf{d}) microring sources, as well as from a model based on the linear resonator characteristics (\textbf{e}).}{fig:spectral}


We pumped on resonance with the cavity at $\nu_p$, and collected signal and idler photons from adjacent cavity resonances, one free spectral range (FSR) over, at $\nu_{s,i} = \nu_p \pm 800\text{ GHz}$. The cavity linewidth was 21~GHz. Source resonances cause the highlighted dips in the transmission spectrum of Fig.~\ref{fig:spectral}a; the peaks in that spectrum are due to the signal-idler filters, discussed later. Our pump laser produced 10.8-ps pulses, with a 40~GHz linewidth, at a rate of 51~MHz. Since SFWM takes in two pump photons, its efficiency scales quadratically with pump power for low squeezing values. However, due to the strong two-photon absorption (TPA) of silicon in the near-infrared, this quadratic scaling only holds at low power, before TPA starts to dominate\cite{Husko2013}. In our measurements, an average pump power of 150~$\mu$W (253~mW peak) was delivered, leading to generation probabilities of 0.06 and 0.09 pairs per pulse for the two sources. System losses reduced these at-source generation rates to around 30 measured coincidences per second, with a coincidence-to-accidental ratio of around 10. An imbalance in source efficiency between the top and bottom was somewhat compensated by the measured 54\% reflectivity of the first coupler, leading to a source balance of $\beta = 43\%$.

Interference between photons from different sources requires those photons to be indistinguishable in all degrees of freedom, but spectral indistinguishability poses a particular challenge. We refer to this spectral indistinguishability as the overlap, $\overlap$, which runs from $\overlap = 0$ for fully distinguishable photon pairs to $\overlap = 1$ for indistinguishable ones. We explored the overlap between the two microring sources by configuring the device to interfere the signal-idler superposition on the `idler' interferometer of Fig.~\ref{fig:schematic}, which was configured as a beamsplitter, and sweeping one source resonance over the other (see Methods). In this way, we could observe two-photon fringes analogous to those in ref.~\citenum{Silverstone2014}, and measure changes in the quality of the quantum interference as the two sources were tuned together. The fringe visibility as a function of source detuning is plotted in Fig.~\ref{fig:spectral}b. Accounting for source imbalance, and multi-pair events, we compute the maximum ($\sigma = 1$) observable visibility as $96.0\%$. We observed a peak visibility of $95.8 \pm 2.1 \%$ when the two sources were tuned. This corresponds to a near-perfect overlap of $\sigma = 0.99 \pm 0.08$. See Supplementary Section~2 for details. When the two sources were completely detuned, the visibility reached a floor of $37\%$, caused by residual interference between broadband photon pairs produced in the non-resonant parts of the interferometer and the single remaining microring source. This visibility indicates that the spectral brightness of the bus waveguide was 1\% of that of the tuned microring. Since all subsequent measurements were performed through the on-chip filters, this waveguide-generated flux did not significantly contribute to our statistics. These data show that the two microring sources could be made indistinguishable, and exhibited brightness which dominated the background SFWM occurring in the rest of the interferometer.

High-purity photon-pair sources---for heralded- or multi-photon experiments---require that, given the frequency of the signal photon, we gain minimal information about the frequency of the idler photon, and vice versa---their frequency states are separable. By pumping the source cavities with spectrally broad pulses, we relax the energy and momentum requirements of the SFWM process. The emitted signal and idler photons then naturally and independently take on the structure of the cavity enhancement, which has been predicted to improve their spectral separability\cite{Helt2010}. To quantify this separability, we measured the signal-idler joint spectral density (JSD) using the stimulated emission tomography method of ref.~\citenum{Liscidini2013}.

Measured JSD profiles are shown in Figs.~\ref{fig:spectral}c and \ref{fig:spectral}d, for the top and bottom microring sources showing an overlap of $\sigma = 0.962$. They exhibit a residual spectral entanglement (and corresponding multi-mode squeezing) with the number of modes quantified by the Schmidt number $K$, where $1/K$ would be the visibility of a triggered Hong-Ou-Mandel interference dip. We measured $K > 1.19$ for the top source, and $K > 1.17$ for the bottom source, where $K = 1$ represents perfect two-mode squeezing, and spectral separability. These values represent lower bounds on $K$ because our measurements only give information on the magnitude of the joint spectral amplitude (JSA), not the phase. We use the bright-light response, obtained by scanning a tuneable laser over the source resonances, to inform our model for the SFWM inside each source resonator. This model takes in the cavity response, the pump lineshape, and the waveguide dispersion and gives as output the theoretical JSD of Fig.~\ref{fig:spectral}e (see Supplementary Section~5 for model details). The theoretically predicted linewidth is somewhat narrower than what we measured; we attribute the difference to self-phase modulation occurring inside each cavity. In straight-waveguide source designs, spectral separability is only achievable by inserting a narrow spectral filter after the pair-generation process\cite{Matsuda2012, Silverstone2014, Harada2011}, which necessarily reduces the source brightness. For bright heralded photon-pair sources, a naturally un-correlated joint spectral density, like those we have shown, is desirable.


Each pair is produced in a superposition of being generated in the top and bottom microring sources simultaneously, since we pump with only enough power to produce one photon pair and there is a fixed phase relationship between the two sources. This pair superposition is then converted into an entanglement between two qubits, each composed of a single signal or idler photon in one or another of two waveguide paths. The signal and idler photons are separated by on-chip filters. These filters, formed by double-bus microring resonators\cite{Rabus2007}, exhibited a selectivity of 22~dB, a bandwidth of 35~GHz, and a loss which was negligible compared to the system loss. Their 640-GHz FSR was designed to select the signal photon, while maximally rejecting the idler (see peaks in Fig.~\ref{fig:spectral}b). Finally, the frequency demultiplexed waveguides are rearranged to group the signal and idler paths together. Written in the form of a density matrix, and in terms of the experimental parameters $\balance$, $\overlap$, and $\Theta$, the resulting qubit-basis state is
\begin{equation}
\begin{split}
	\theorystate \ =\ \ &\ket{{00}}\bra{{00}} \cdot \balance \\
		+\ &\ket{{11}}\bra{{11}} \cdot (1-\balance)  \\
		+\ &\ket{{00}}\bra{{11}} \cdot e^{-i\Theta}\sqrt{\balance}\sqrt{1-\balance} \cdot \overlap\\
		+\ &\ket{{11}}\bra{{00}} \cdot e^{+i\Theta}\sqrt{\balance}\sqrt{1-\balance} \cdot \overlap^*
\end{split}
\label{eq:state}
\end{equation}
where the balance $\balance$ describes the relative brightness of the two sources, the overlap $\overlap$ quantifies the spectral indistinguishability of the two sources, and $\Theta$ accumulates the intrinsic total phase between the two qubits. We define a photon in the top (bottom) waveguide of each qubit to be a logical \ket{0} (\ket{1}), and the first (second) qubit of each pair to be the signal (idler). For example, $\ket{00}\bra{00}\balance$ means both qubits have a photon in the top mode with probability $\balance$. Experimentally, we can control the balance $\balance$ by adjusting the tuning of the filters (at the expense of spectral overlap), and we control the overlap $\overlap$ by tuning the two microring sources. If the flux from the two sources is balanced ($\balance = 1/2$) and the source joint spectra overlap perfectly ($\overlap = 1$) then $\theorystate$ is in the family of maximally entangled Bell states. If $\balance \in \{0,1\}$ then $\theorystate$ is separable; if $\overlap = 0$ then $\theorystate$ is mixed. See Supplementary Section~1 for a detailed state evolution, and a short discussion on the origin of the entanglement in this device.


The on-chip state was manipulated and measured using integrated single-qubit analysis Mach-Zehnder interferometers. These interferometers, shown in Fig.~\ref{fig:schematic}, implemented $\Rz$ and $\Ry$ rotations by angles $\thSZ$,  $\thIZ$, $\thSY$,  and $\thIY$ on the signal (S) and idler (I) qubits, respectively. These rotations facilitated single-qubit measurements on the generated two-qubit state. Photons from the two qubits were counted using coincidences between two 25\%-efficient avalanche photodiodes, gated on each laser pulse.


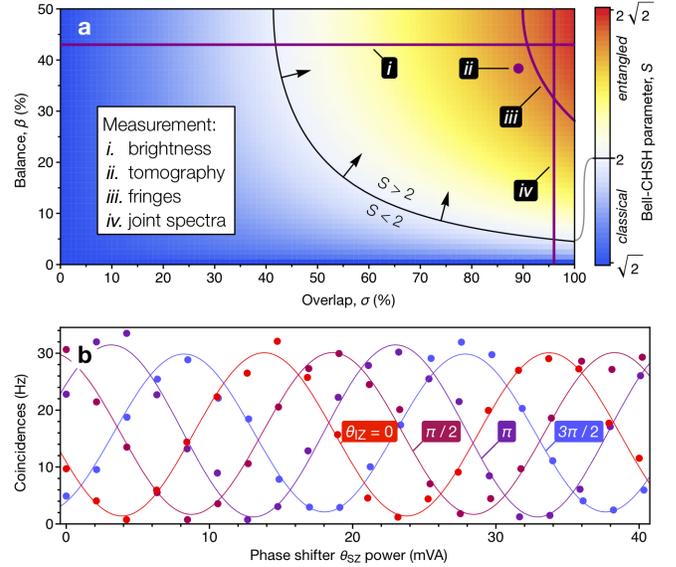
\makefigc{fchsh}{Summary of measurements in the context of Bell-CHSH inequality violation. \textbf{a,} Map showing violation $S$ as a function of source balance $\balance$ and overlap $\overlap$, with listing of measurement results overlaid. If $S<2$, measurement correlations on the state can be described classically, while if $S>2$, a quantum description is required as the state is entangled. By measuring: (i) the brightness of each source, we can estimate the balance $\balance$; (ii) the quantum state, via quantum state tomography we can estimate both the balance $\balance$ and the overlap $\overlap$; (iii) correlated fringes we obtain a value for the violation $S(\balance,\overlap)$; and (iv) the overlap between measured joint spectra gives $\overlap$. The measurement of $\overlap$ in (iv) naturally excludes multi-photon contamination, while the other measurements (i-iii) necessarily include it, and result in lower values of $\overlap$ as a consequence. \textbf{b,} Fringes generated by $\Rz$ rotations on signal and idler qubits, allowing a direct measurement of CHSH $S$ parameter (denoted measurement (iii) in part a). }{fig:chsh}

A well known test of quantum non-locality, as well as an indicator of the entanglement present in a quantum state\cite{EntanglementMeasexp2013}, is based on the reformulation of Bell's original inequality due to Clauser, Horne, Shimony and Holt (CHSH)\cite{CHSH1969}. In this test, a parameter $S$ is evaluated\cite{Aspect1982}, whose value indicates the presence of non-locality: if $S>2$, the state is non-local; if $S=2\sqrt 2$ the state is maximally entangled.

We can explicitly calculate the $S$ which results from $\theorystate$ of Eq.~\ref{eq:state}, to quantify how the violation depends on the balance and source overlap (see Supplementary Section~3):
\begin{equation}
	S = \sqrt{2} \(1 + 2 \overlap \sqrt{\balance}\sqrt{1 - \balance} \)
	\label{eq:modelchsh}
\end{equation}
which reaches the maximum violation of $2\sqrt{2}$ when $\overlap = 1$ and $\balance = 1/2$, and decreases as the overlap and balance deviate from these values. Equation~\ref{eq:modelchsh} is plotted as a contour in Fig.~\ref{fig:chsh}a, showing the level of entanglement indicated by each of our measurements ($i$-$iv$).

One manifestation of the entanglement present in our on-chip state (equation~\ref{eq:state}) is the presence of the non-local phase factor~$\Theta$. As a result of this factor, $\Rz$ rotations applied to each qubit cannot be observed independently: each equally contributes to the total phase of the state, $\Theta$. To demonstrate this, we configured the signal and idler $\Ry$ rotations to mix the two modes of each qubit ($\thSY = \thIY = \pi/2$), then manipulated both $\thSZ$ and $\thIZ$, and observed coincidence fringes with an entangled phase $\Theta = \thSZ + \thIZ$, shown in Fig.~\ref{fig:chsh}b. These fringes exhibit a mean visibility of $94.7\pm1.0\%$ which is consistent with a strong CHSH violation of $S = 2.686\pm 0.026$. This value of $S$ violates the inequality by 83\% and by 26 standard deviations.


For any quantum system the total accessible information of its quantum state  is encoded in its density matrix $\expstate$. Quantum state tomography\cite{Fano1957, Leonhardt1995} is the process of experimentally estimating $\expstate$ based on a series of measurements. We made an over-complete set\cite{Measurementsofqubits} of thirty-six measurements on the state using the on-chip interferometers, and used the results to estimate $\expstate$. See Supplementary Section~4 and Methods for details. We produced a series of on-chip states---those that were separable, mixed, and entangled---by using different configurations of source and filter tuning. Manipulated both the source balance ($\balance$) and frequency overlap ($\overlap$), we observed changes in the resulting state in agreement with the predictions of equation~\ref{eq:state}. In the first measurement, we tuned only the top source and filter, and detuned the bottom filter, effectively setting $\balance = 1$. We estimated the state shown in Fig.~\ref{fig:tomography}a, which exhibits a peak in the pure \ket{{00}} component, as expected (a similar result was obtained with the top filter detuned, with $96\pm1\%$ fidelity, see Supplementary Figure~6). Next, we tuned both filters to match each source, but did not tune the two sources to overlap, effectively setting $\balance=1/2$ and $\overlap = 0$. We observed amplitude in both the \ket{{00}} and \ket{{11}} components, but without coherence terms ($\ket{{00}}\bra{{11}}$ and $\ket{{11}}\bra{{00}}$), as shown in Fig.~\ref{fig:tomography}b. As predicted by equation~\ref{eq:state}, this is due to a lack of spectral overlap between photons produced in the top and bottom sources, which results in no interference at the analysis interferometers. Indeed, the estimated state is mixed, with a purity of $0.49\pm0.01$ (with 0.5 expected). Due to the lack of coherence, we were able to use the filter lineshapes to balance the source brightness, achieving $\balance = 0.49$. Finally, we tuned all four microrings to overlap, and measured the highly entangled state of Fig.~\ref{fig:tomography}c, in which both sources are producing photons, and are mutually coherent.

We evaluated the Bell-CHSH $S$ parameter for the above entangled state (Fig.~\ref{fig:tomography}c), and found $S = 2.692\pm 0.018$. This value violates the inequality by 83\% and by 38 standard deviations, and is in excellent agreement with our estimation based on correlated fringes (Fig.~\ref{fig:chsh}b).

To compare the probability amplitudes of a measured state $\expstate$ with those of an expected state $\theorystate$, we evaluate the fidelity $F$ as
\begin{equation}
F = \text{Tr}\mkern-2mu\(\sqrt{\sqrt{\theorystate} \cdot \expstate \cdot \sqrt{\theorystate}}\)
\end{equation}
where the matrix square root is defined as $\sqrt{\hat\rho}\cdot\sqrt{\hat\rho} = \hat\rho$. The fidelity $F$ of two states runs from 0 to 1: $F=1$ indicates the states are identical, while $F=0$ indicates orthogonality. We used $F$ to gauge the ability of our device to prepare the three target states indicated in Figs.~\ref{fig:tomography}a-c, finding $F > 90\%$ in all cases.

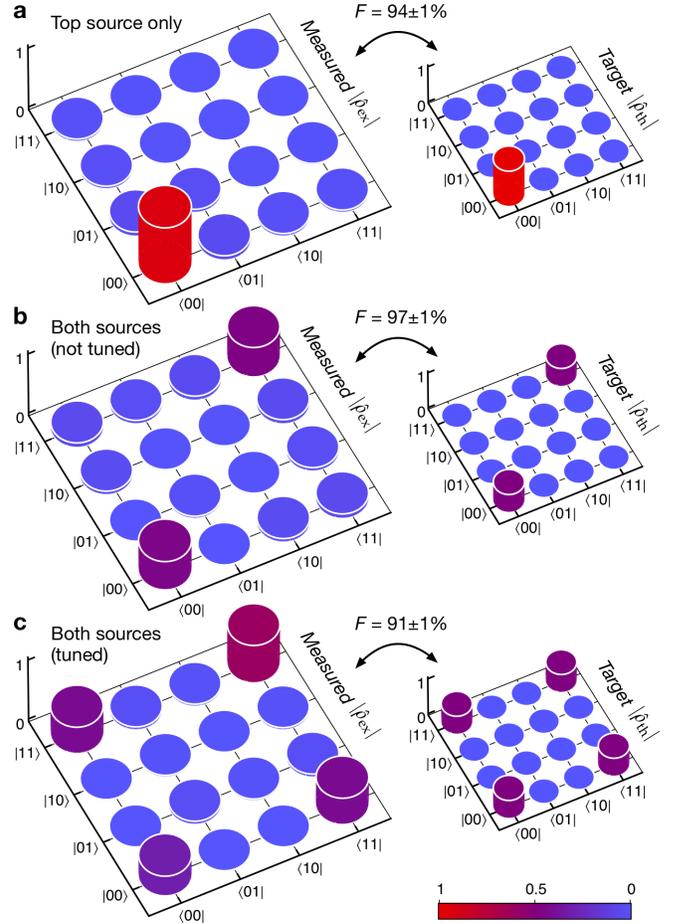
\makefigc{ftomo}{On-chip states for various device configurations, estimated using integrated analysis interferometers. Measured states are enlarged at left, with target states and corresponding fidelity (as defined in text) at right. State corresponding to \textbf{a,} top source only (with bottom source detuned) \textbf{b,} both sources tuned but not overlapped, showing mixed state, and \textbf{c,} both sources tuned and overlapped, showing path qubit entanglement.}{fig:tomography}


We have demonstrated bright and spectrally-separable photon-pair sources, phase-stable frequency-selective elements, and passive and active optics integrated on a silicon chip, and used them together to generate and analyse path-qubit entanglement at optical frequencies compatible with telecommunications networks. We used a new method\cite{Liscidini2013} to provide evidence that the silicon microring source can produce spectrally uncorrelated photons---making this structure a promising candidate for future multi-pair experiments on silicon chips. Moreover, we were able to overlap two such microrings to a high degree, obtaining high-visibility quantum interference between them, despite their well-documented\cite{Pernice2011, Pelc2014} nonlinear dynamics. We showed how the on-chip state strongly violates the Bell-CHSH test---a strict test of entanglement---and confirmed this experimentally in several different ways, including via an on-chip quantum state tomography. That such a high degree of entanglement is generated and preserved by the device indicates the high-fidelity operation of all its constituent parts.

By assembling and characterising a path-entangled Bell state on-chip, we have shown that silicon photonics, with its inherently mature and scalable manufacturing process, can be used to produce the most basic cluster state---one formed of two qubits. Large cluster states, combined with the single-qubit measurements we have shown here, are the resources needed for measurement-based quantum computation, a promising approach to universal quantum computation with photons\cite{Briegel2009, Nielsen2004, OBrien2009}. Producing these exotic states remains a great challenge, but our results show that microring SFWM sources and on-chip frequency manipulation will be useful tools for producing the indistinguishable and frequency-separable photons which are needed to engineer photonic quantum information systems to scales well beyond today's experiments.

\section*{Methods}
\footnotesize

\pseudosection{Device fabrication.} The device was fabricated on a silicon-on-insulator wafer with a 220-nm silicon slab and a 2-$\mu$m buried oxide layer. The  waveguides were 500~nm wide, and were patterned using direct-write electron beam lithography into a hydrogen silsesquioxane resist layer, used as a hard mask for the reactive ion etching of the silicon slab. These structures were subsequently coated with a 900-nm silica layer. Phase shifters were based on resistive heaters, patterned atop the silica layer using a lift-off technique on a 50-nm nickel-chromium film. Electrical traces connecting the heater elements were similarly patterned in a 200-nm gold layer.

All waveguide-waveguide couplers were fabricated as evanescent field (directional) couplers with 300-nm gaps. Losses were minimised at waveguide crossings via tapered sections, and fibre-to-chip coupling was achieved using inverse silicon tapers embedded in $2\times1.5$-$\mu$m$^2$ SU8 polymer waveguides.

\pseudosection{Source overlap measurement.} To obtain the data of Fig.~\ref{fig:spectral}b, we spectrally swept the top microring source resonance across the bottom one, while interfering the generated pairs on the bottom Mach-Zehnder interferometer (MZI, denoted $\Ry(\thIY)$ in Fig.~\ref{fig:schematic}). To allow both signal and idler photons to reach the bottom MZI, we detuned both filter microrings, such that they were effectively removed from the light path, and both signal and idler photons were reflected downwards. To allow interference to occur on the bottom MZI, we configured it as a simple beamsplitter by setting $\thIY = \pi/2$. We then measured coincidences across the bottom two output ports of the device (labelled $\ket{0}_i$ and $\ket{1}_i$ in Fig.~\ref{fig:schematic}) while at the same time varying $\thIZ$ to form fringes. We fit these fringes sinusoidally to extract the visibility of each, and these visibility data are plotted in Fig.~\ref{fig:spectral}b.

\pseudosection{Projector calibration.} The rotations $\Ry(\thSY)$, $\Ry(\thIY)$, $\Rz(\thSZ)$, and $\Rz(\thIZ)$ were used to analyse the states generated on-chip. To do so, we calibrated the phase-voltage relationship of each phase shifter independently. We injected laser light into the device and recorded the output intensity $I$ from each interferometer as a function of the phases, obtaining $I(\thSY, \thSZ)$ and $I(\thIY, \thIZ)$. We fit the data with a model of the double interferometer (which included the first on-chip coupler), yielding the various coupler reflectivities and phase-voltage relationships. These data and models are plotted in Supplementary Figure~5. Since we were unable to determine the absolute values of $\thSZ$ and $\thIZ$, we defined these phases relatively. We used the resulting models to control the on-chip phase shifters as required by each part of the experiment.

\pseudosection{Quantum state tomography.} We used the on-chip rotations to implement an informationally over-complete\cite{OverComTom} set of thirty-six projective measurements on two qubits, to reconstruct $\expstate$. Each measurement projected each of the two qubits onto one of the six states: $\ket{0},\ket{1},\ket{+},\ket{-},\ket{{+i}}$ or $\ket{{-i}}$. We then performed a multi-dimensional search for the two-qubit state $\theorystate$ which could best explain the measurement outcomes, based on a constrained least squares (CLS) estimator. The problem is defined as:
\begin{equation}
	\theorystate = \underset { \hat\rho  }{ \mathrm{arg\, min} } \left\{\underset {i}{\sum}\left|P_{\mathrm{ex}}(i)-P_{\hat\rho}(i)\right|^2\right\}
	\label{eq:CLS}
\end{equation}
Where $\hat\rho$ is the density matrix generated internally by the search algorithm, with the condition that it is physical: i.e. hermitian, positive semi-definite, and with trace one. $P_{\mathrm{ex}}(i)$ is the $i^\mathrm{th}$ experimental probability estimate, and $P_{\hat\rho}(i)$ is the corresponding computed result based on the application of the $i^\text{th}$ projector to $\hat\rho$.

Experimental uncertainty per count was measured using residuals from a number of coincidence fringes. This was used to estimate the uncertainty on each $P_\mathrm{ex}(i)$. A Monte-Carlo method was then used to sample 500 reconstructions around the measured values $P_\mathrm{ex}(i)$, and the uncertainty in each tomographic parameter (fidelity, $S$, etc.) was estimated from the distribution of these reconstructions.

\pseudosection{Joint spectral density measurement.} In obtaining the data of Figs.~\ref{fig:spectral}c and \ref{fig:spectral}c, we followed closely the prescription of Liscidini et al.\cite{Liscidini2013}, and the method of Eckstein et al.\cite{Eckstein2014}. We tuned each ring separately, and pumped them as detailed in the main text. A narrow linewidth seed laser was swept across one resonance of each ring, and the stimulated four-wave mixing (FWM) was collected by a spectrometer. The seed field was provided by an amplified tuneable laser with 10~kHz linewidth (Photonetics Tunics Plus). We reduced the launched seed power until no evidence of seed-induced optical bi-stability remained. The stimulated FWM signal was collected by an optical spectrum analyser with a 6-GHz resolution (Anritsu MS9740A).

\bibliographystyle{unsrt}

\section*{Acknowledgements}
We thank John G. Rarity, Peter J. Shadbolt, and Anthony Laing for valuable discussions, as well as Luka Milic for his help with the joint spectral density measurements. We also thank the staff of both the James Watt Nanofabrication Centre in Glasgow, and the Centre for Nanoscience and Quantum Information in Bristol for their support. We acknowledge support from the European Research Council through the BBOI project. M.G.T. acknowledges support from an Engineering and Physical Sciences Research Council (EPSRC, UK) Early Career Fellowship. J.W.S. acknowledges an EPSRC Doctoral Training Account, and a Natural Sciences and Engineering Research Council (Canada) Canada Graduate Scholarship. J.L.O'B. acknowledges a Royal Society Wolfson Merit Award and a Royal Academy of Engineering Chair in Emerging Technologies.


\begin{thebibliography}{10}

\bibitem{Raussendorf2003}
R.~Raussendorf and H.~J. Briegel.
\newblock {A one-way quantum computer}.
\newblock {\em Phys. Rev. Let.}, 86(22):5188, 2001.

\bibitem{Horodecki2009}
R.~Horodecki, M.~Horodecki, and K.~Horodecki.
\newblock {Quantum entanglement}.
\newblock {\em Rev. Mod. Phys.}, 81(2):865--942, 2009.

\bibitem{Briegel2009}
H.~J. Briegel, D.~E. Browne, W.~Duer, R.~Raussendorf, and M.~Van~den Nest.
\newblock {Measurement-based quantum computation}.
\newblock {\em Nature Phys.}, 5(1):19--26, 2009.

\bibitem{Nielsen2004}
M.~A. Nielsen.
\newblock {Optical Quantum Computation Using Cluster States}.
\newblock {\em Phys. Rev. Lett.}, 93(4):040503, 2004.

\bibitem{OBrien2009}
J.~L. O'Brien, A.~Furusawa, and J.~Vuckovic.
\newblock {Photonic quantum technologies}.
\newblock {\em Nature Photon.}, 3(12):687--695, 2009.

\bibitem{Pernice2011}
W.~H.~P. Pernice, C.~Schuck, M.~Li, and H.~X. Tang.
\newblock Carrier and thermal dynamics of silicon photonic resonators at
  cryogenic temperatures.
\newblock {\em Opt. Express}, 19(4):3290--3296, 2011.

\bibitem{Pernice2010}
W.~H.~P. Pernice, M.~Li, and H.~X. Tang.
\newblock {Time-domain measurement of optical transport in silicon micro-ring
  resonators}.
\newblock {\em Opt. Express}, 18(17):18438--18452, 2010.

\bibitem{Priem2005}
G.~Priem, D.~Van Thourhout, P.~Dumon, W.~Bogaerts, G.~Morthier, and R.~Baets.
\newblock {Optical bistability and pulsating behaviour in Silicon-On-Insulator
  ring resonator structures}.
\newblock {\em Opt. Express}, 13(23):9623--9628, 2005.

\bibitem{Helt2010}
L.~G. Helt, J.~E. Sipe, Z.~Yang, and M.~Liscidini.
\newblock {Spontaneous four-wave mixing in microring resonators.}
\newblock {\em Opt. Lett.}, 35(18):3006--3008, 2010.

\bibitem{Aspect1982}
A.~Aspect, P.~Grangier, and G.~Roger.
\newblock Experimental realization of einstein-podolsky-rosen-bohm
  gedankenexperiment: A new violation of bell's inequalities.
\newblock {\em Phys. Rev. Lett.}, 49:91--94, 1982.

\bibitem{Yao2012}
Xing-Can Yao, Tian-Xiong Wang, Ping Xu, He~Lu, Ge-Sheng Pan, Xiao-Hui Bao,
  Cheng-Zhi Peng, Chao-Yang Lu, Yu-Ao Chen, and Jian-Wei Pan.
\newblock {Observation of eight-photon entanglement}.
\newblock {\em Nature Photon.}, 6(4):225--228, 2012.

\bibitem{Bell2013}
B~A Bell, M~S Tame, A~S Clark, WJ~Wadsworth, R~W Nock, and J~G Rarity.
\newblock {Experimental characterization of universal one-way quantum
  computing}.
\newblock {\em New J. Phys.}, 15(5):053030, 2013.

\bibitem{Shadbolt2012}
Peter~J Shadbolt, M~R Verde, A~Peruzzo, A~Politi, A~Laing, M~Lobino, Jonathan
  C~F Matthews, Mark~G Thompson, and J~L O'Brien.
\newblock {Generating, manipulating and measuring entanglement and mixture with
  a reconfigurable photonic circuit}.
\newblock {\em Nature Photon.}, 6(1):45--49, 2012.

\bibitem{Politi2008}
A~Politi, M~J Cryan, J~G Rarity, S~Yu, and J~L O'Brien.
\newblock {Silica-on-silicon waveguide quantum circuits}.
\newblock {\em Science}, 320:646, 2008.

\bibitem{Silverstone2014}
J.~W. Silverstone, D.~Bonneau, K.~Ohira, N.~Suzuki, H.~Yoshida, N.~Iizuka,
  M.~Ezaki, C.~M. Natarajan, M.~G. Tanner, R.~H. Hadfield, V.~Zwiller, G.~D.
  Marshall, J.~G. Rarity, J.~L. O'Brien, and M.~G. Thompson.
\newblock On-chip quantum interference between silicon photon-pair sources.
\newblock {\em Nature Photon.}, 8(2):104--108, 2014.

\bibitem{Collins2013}
Matthew~J Collins, Chunle Xiong, Trung~D Vo, Jiakun He, Shayan Shahnia,
  Christopher Reardon, M~J Steel, Thomas~F Krauss, Alex~S Clark, and Benjamin~J
  Eggleton.
\newblock {Integrated spatial multiplexing of heralded single photon sources}.
\newblock {\em Nat. Commun.}

\bibitem{Sharping2006}
J.~E. Sharping, K.~F. Lee, M.~A. Foster, A.~C. Turner, B.~S. Schmidt,
  M.~Lipson, A.~L. Gaeta, and P.~Kumar.
\newblock {Generation of correlated photons in nanoscale silicon waveguides.}
\newblock {\em Opt. Express}, 14(25):12388--12393, 2006.

\bibitem{Agrawal2006}
Q.~Lin and G.~P. Agrawal.
\newblock Silicon waveguides for creating quantum-correlated photon pairs.
\newblock {\em Opt. Lett.}, 31(21):3140--3142, 2006.

\bibitem{Matsuda2012}
N.~Matsuda, H.~Le~Jeannic, H.~T. Fukuda, W.~J. Munro, K.~Shimizu, K.~Yamada,
  Y.~Tokura, and H.~Takesue.
\newblock A monolithically integrated polarization entangled photon pair source
  on a silicon chip.
\newblock {\em Sci. Rep.}, 2:817, 2012.

\bibitem{Crespi2011}
A.~Crespi, R.~Ramponi, R.~Osellame, L.~Sansoni, I.~Bongioanni, F.~Sciarrino,
  G.~Vallone, and P.~Mataloni.
\newblock {Integrated photonic quantum gates for polarization qubits}.
\newblock {\em Nat. Commun.}, 2:566, 2011.

\bibitem{Olislager2013}
L.~Olislager, J.~Safioui, S.~Clemmen, K.~P. Huy, W.~Bogaerts, R.~Baets,
  P.~Emplit, and S.~Massar.
\newblock {Silicon-on-insulator integrated source of polarization-entangled
  photons}.
\newblock {\em Opt. Lett.}, 38(11):1960--1962, 2013.

\bibitem{Azzini2012}
S.~Azzini, D.~Grassani, M.~J. Strain, M.~Sorel, L.~G. Helt, J.~E. Sipe,
  M.~Liscidini, M.~Galli, and D.~Bajoni.
\newblock Ultra-low power generation of twin photons in a compact silicon ring
  resonator.
\newblock {\em Opt. Express}, 20:23100--23107, 2012.

\bibitem{Husko2013}
C~A Husko, A~S Clark, M~J Collins, and A~De~Rossi.
\newblock {Multi-photon absorption limits to heralded single photon sources}.
\newblock {\em Sci. Rep.}, 3:3087, 2013.

\bibitem{Liscidini2013}
M.~Liscidini and J.~E. Sipe.
\newblock {Stimulated Emission Tomography}.
\newblock {\em Phys. Rev. Lett.}, 111(19):193602, 2013.

\bibitem{Harada2011}
K.~Harada, H.~Takesue, H.~Fukuda, T.~Tsuchizawa, T.~Watanabe, K.~Yamada,
  Y.~Tokura, and S.~Itabashi.
\newblock {Indistinguishable photon pair generation using two independent
  silicon wire waveguides}.
\newblock {\em New J. Phys.}, 13(6):065005, 2011.

\bibitem{Rabus2007}
D.~G. Rabus.
\newblock {\em {Integrated Ring Resonators}}.
\newblock Springer, Berlin Heidelberg New York, 2007.

\bibitem{EntanglementMeasexp2013}
K.~Bartkiewicz, B.~Horst, K.~Lemr, and A.~Miranowicz.
\newblock Entanglement estimation from bell inequality violation.
\newblock {\em Phys. Rev. A}, 88:052105, 2013.

\bibitem{CHSH1969}
J.~F. Clauser, M.~A. Horne, A.~Shimony, and R.~A. Holt.
\newblock Proposed experiment to test local hidden-variable theories.
\newblock {\em Phys. Rev. Lett.}, 23:880--884, 1969.

\bibitem{Fano1957}
U.~Fano.
\newblock Description of states in quantum mechanics by density matrix and
  operator techniques.
\newblock {\em Rev. Mod. Phys.}, 29:74--93, 1957.

\bibitem{Leonhardt1995}
U~Leonhardt.
\newblock Quantum-state tomography and discrete wigner function.
\newblock {\em Phys. Rev. Lett.}, 74:4101--4105, 1995.

\bibitem{Measurementsofqubits}
D.~F.~V. James, P.~G. Kwiat, W.~J. Munro, and A.~G. White.
\newblock Measurement of qubits.
\newblock {\em Phys. Rev. A}, 64:052312, 2001.

\bibitem{Pelc2014}
J.~S. Pelc, K.~Rivoire, S.~Vo, C.~Santori, D.~A. Fattal, and R.~G. Beausoleil.
\newblock Picosecond all-optical switching in hydrogenated amorphous silicon
  microring resonators.
\newblock {\em Opt. Express}, 2013.

\bibitem{OverComTom}
M.~D. de~Burgh, N.~K. Langford, A.~C. Doherty, and A.~Gilchrist.
\newblock Choice of measurement sets in qubit tomography.
\newblock {\em Phys. Rev. A}, 78:052122, 2008.

\bibitem{Eckstein2014}
A.~Eckstein, G.~Boucher, A.~Lema{\^\i}tre, P.~Filloux, I.~Favero, G.~Leo,
  M.~Liscidini, and S.~Ducci.
\newblock High-resolution spectral characterization of two photon states via
  classical measurements.
\newblock {\em Laser {\&} Phot. Rev.}, 8(5):L76--L80, 2014.

\end{thebibliography}

\end{document}